\documentstyle[12pt]{article}

\newcommand{\be}{\begin{equation}}
\newcommand{\ee}{\end{equation}}

\newcommand{\bea}{\begin{eqnarray}}
\newcommand{\eea}{\end{eqnarray}}

\newcommand{\p}{\partial}

\newcommand{\nn}{\nonumber \\}
\newcommand{\f}{\frac}
\newcommand{\w}{\wedge}

\begin{document}
\setcounter{page}{0}
\thispagestyle{empty}

\begin{flushright}
{\bf arXiv: 0904.3620}
\end{flushright}
\begin{center} \noindent \Large \bf
Non-relativistic supersymmetric Dp branes
\end{center}

\bigskip\bigskip\bigskip
\vskip 0.5cm
\begin{center}
{ \normalsize \bf  Shesansu Sekhar Pal}

\vskip 0.5cm

\vskip 0.5 cm
Saha Institute of Nuclear Physics, \\
1/AF, Bidhannagar, Kolkata 700 064, India\\
\vskip 0.5 cm
\sf shesansu${\frame{\shortstack{AT}}}$gmail.com
\end{center}
\centerline{\bf \small Abstract}

The supergravity solutions to various $p$-branes are presented and typically 
it breaks one quarter of   supersymmetry except for the D1 
brane case, where it breaks  one half  of supersymmetry and are manifestly  
non-relativistic in nature. The symmetries that the solutions enjoys 
are that of space and time translations, 
rotations, boosts but without any scaling and special conformal transformations
except for $p\neq 3$. We have also constructed supersymmetric 
 non-relativistic cascading 
solutions to intersecting $D3$ and $D5$ branes on both the singular as well as
on deformed conifold, where the $D5$'s are wrapped on the $S^2$ of the 
Calabi-Yau i.e. the analogs of Klebanov-Tseytlin and Klebanov-Strassler 
solutions and the supersymmetric  non-relativistic M2 brane solution.

\newpage
\section{Introduction}

Recently, the  gauge-gravity duality has been proposed  to understand the
 strongly coupled behavior of a very specific field theory \cite{jm},
\cite{gkp}, \cite{ew}, which is reviewed in \cite{agmoo} and is further  
reviewed for non-conformal field theories in \cite{oa}.  It is plausible
 that the same  prescription may even 
hold to understand the behavior of strongly coupled electrons for example 
in condensed matter physics. 
 Now,  
in order to apply it, we need to construct a dual gravitational 
 system that possesses the required symmetries. For the present case the 
symmetries 
are that of  the non-relativistic symmetries i.e. the Schrodinger 
symmetry. The generators associated to  such symmetries are that of 
translational invariance of both space and time coordinates, rotation, boosts,
scaling invariance and a special conformal invariance. 
In this context, there have been some proposals put forward and
 one such example, with  the necessary symmetries, 
 for which the metric takes the following form
\be\label{bmz}
ds^2=-r^{2z}(dx^+)^2+r^2(-2dx^+dx^-+dx^2_i)+\f{dr^2}{r^2},
\ee 
 was suggested in \cite{son} and \cite{bm}. It is 
interesting to note that this particular choice to metric preserves all the 
symmetries of Schrodinger group but only when the exponent, $z$, takes a 
specific value that is for $z=2$. Moreover, it is suggested that when the 
exponent takes such a value, it is dual to systems of cold atoms. 
Away from this value of the exponent, 
there won't be any special conformal symmetry, however, the other generators 
of the Schrodinger group will be there for  $z\neq 2$. 
This solution 
has been successfully  embedded  in  10 dimensional string theory  in 
\cite{hrr},~\cite{mmt} and \cite{abm} but only for $z=2$. It is important
to emphasize that, unfortunately these  extremal 
solution do not preserve any supersymmetry and hence it is not a priori clear 
the stability of such  solutions. In earlier studies, \cite{hkss} and 
\cite{hkms}  have made attempts  to  construct and understand the 
gravitational systems dual to  quantum critical points, 
which obeys the relativistic conformal symmetries.
 Based on such suggestions, generically we may expect 
that such solutions should preserve some amount of supersymmetry at the 
critical point from the stability point of view.

 In \cite{hy}, it is reported that  there exists solutions 
that preserves
 different amount  of  supersymmetry depending on the choice of the manifold
i.e. the choice of the  5-dimensional metric in the direction perpendicular to 
D3 brane world volume. 
The explicit form of the solution (in their notation)
\bea
ds^2&=&r^2(-2dx^+dx^-+dx^2_i)-r^{2z}f(X_5)(dx^+)^2+\f{dr^2}{r^2}+ds^2(X_5),\nn
F_5&=& 4(1+\star_{10})vol(X_5)
\eea
where $f(X_5)$ is an eigenfunction  of the Laplacian constructed on the 
5-dimensional manifold $X_5$ and required to  obey
\be
-\nabla^2_{X_5}f=4(z^2-1)f.
\ee
For $z=2$, the D3 brane solution
preserves $1/4$ supersymmetry for $X_5=S^5$ and when $X_5$ is any other 
Sasaki-Einstein spaces like $T^{1,1},~Y^{p,q}$ or $L^{p,q,r}$ etc., it 
 preserves $1/16$ supersymmetries.

It is interesting to note that this is not\footnote{For $z=2$, the only known 
supersymmetric D3 brane solution is given in \cite{hy}.} the only kind of 
supersymmetry preserving D3 brane solutions for $z>2$, there also exists 
another class of solution  with   the structure \cite{dg}
\be\label{gen_metric}
ds^2=r^2(2dx^+dx^-+dx^2_i)+2r^2dx^+[C+hdx^+]+\f{dr^2}{r^2}+ds^2(X_5),
\ee 
where both the 1-form  $C$  and the function $h$ are defined on the $CY_3$,  
with the metric $ds^2=dr^2+r^2 ds^2(SE_5)$, 
and are expected to obey
\be\label{harmonic_condition}
d \star_{CY_3}dC=0,~~~ \nabla^2_{CY_3}h=0.
\ee  
These objects are taken  as
\be
C=r^{\lambda}\beta,~~~h=r^{\lambda'} f'
\ee
such that, the 1-form $\beta$ and the function $f'$ obeys 
\be
\Delta_{SE_5}\beta=\mu\beta,~~~d^{\dagger}\beta=0,~~~
-\nabla^2_{SE_5}f'=k'f',
\ee
with the eigenvalue 
\be
\mu=\lambda(\lambda+2),~~~k'=\lambda'(\lambda'+4).
\ee

In fact when $z=4$ and $h=0$, this solution has been originally presented in 
\cite{mmt}. This way of generating solutions has been generalized to 
Calabi-Yau 4-folds as well \cite{dg}, in particular for the $M2$ branes. 

In another context several gravitational descriptions has been given which 
do admit the non-relativistic symmetry group but without the special 
conformal generator and boost generator in \cite{klm} and this is being 
interpreted as systems exhibiting Lifshitz-like behavior. This has been 
generalized  to any arbitrary but even dimensional spacetime \cite{ssp1} 
using a better choice of coordinate system \cite{ssp} and in 
\cite{ssp2}, the gravitational system has been constructed with two explicit 
dynamical exponents, in \cite{amsv} the $1/N$ behavior of \cite{klm} is studied
and in \cite{dt}, the finite temperature behavior of it. The thermodynamic
properties has been studied in \cite{mt} and the N-point 
correlation functions \cite{vw} of eq(\ref{bmz}). 

The other properties of the Galilean algebra and its supersymmetric version 
has been studied in \cite{gca} and the effect of Lorentz violations in 
quantum field theory in \cite{da}.

In the application of solution generating techniques like NMT \cite{ghhlr},
\cite{ag} 
and TsT\cite{mmt}  have been applied to several systems like that of 
Sakai-Sugimoto model in \cite{ssp3} and some more systems in 
\cite{mot}, \cite{ms} and \cite{bk}. While applying the solution 
generating technique 
in particular, NMT to Type IIA theories, we generate solutions where the 
non-compact 
part of the metric and that of the form fields depends on the trigonometric 
functions \cite{ssp3} and \cite{mot} and because of this the separation of 
variables in the equations of motion to the minimally coupled scalar field
 is not possible.  This is mainly due to the 
the presence of spheres of even dimensionality and such spheres 
 cannot be written as a fibration over a complex projective space. However, 
we can avoid the appearance of the trigonometric functions in the metric 
components along the non-compact directions as well as in the form fields 
by writing down topologically, the space transverse to the brane directions as 
$R^1\times S^1\times S^{2n+1}$ or $R^1\times S^{2n+1}\times R^1$. 
As an example, for $D4$ branes we can write
the metric along the directions perpendicular to brane as 
\be\label{IIA}
ds^2=dr^2+r^2d\Omega^2_3+dw^2, 
\ee
instead of $ds^2=dr^2+r^2d\Omega^2_4$, 
where the $S^1$ is defined by $w$ and  has the periodicity $w=w+2\pi$. 
Taking such a choice of topology makes a change in the power of the radial 
coordinate $r$ that appear in  the harmonic function.

In this paper, we shall generalize the
construction of \cite{mmt} and \cite{dg} to construct spacetimes that do not 
admit a Calabi-Yau 3-fold, in particular, we
shall generate supersymmetric solutions for coincident $Dp$ branes for which 
$p$ takes value $1,2,3,4$ and $5$. Moreover, we shall write down the 
precise form of the 1-form $C$, for each case.

This will be further generalized to  construct solutions  
 for which the Calabi-Yau is not any more singular, i.e. which is 
not of the following form 
\be
ds^2(CY_3)=dr^2+r^2ds^2(SE_5).
\ee
 As a  specific example we shall construct solutions on 
 the deformed conifold, where we shall take a bunch coincident $D3$ branes 
extended along the non compact directions and another bunch of coincident 
$D5$ branes that are wrapped on the $S^2$ of the deformed conifold and extended
along the four non-compact directions. Even though the exact 
form of such one form $C$ is non-trivial to find, however, for this particular 
example we have got the solution to  eq(\ref{harmonic_condition}).

In generating solutions as stated in the previous paragraph, we shall assume
 that the function $h$, that  appear in \cite{dg} takes a simple value, $h=0$ 
 and shall construct the 1-form 
$C$, for which it obeys the condition \be d \star_{9-p}dC=0 \ee for both 
Type IIA/IIB theories. The solutions for the examples that we have studied 
preserve $1/4$ supersymmetry. It is interesting to note that the 
supersymmetry preserving criteria do not fixes either the form of $C$ or 
$h$, whose structure  has to be fixed only by solving the equations of motion 
to fields.

When the spacetime do admit a singular 
Calabi-yau 3-fold, one can define a 1-form $C$, following \cite{dg}
\be
C_i={J_i}^j\p_j\Phi,
\ee
where $J$ is the Kahler form of the CY$_3$ and the function $\Phi$ is defined 
on the $CY_3$. Imposition of the condition eq(\ref{harmonic_condition}) 
results in 
\be\label{phi_alpha}
\nabla^2_{CY}\Phi=\alpha,
\ee
where $\alpha$ is a constant. Now, let us assume that  for other cases 
where the solution do not admits a $CY_3$, for example  that in the 
Dp brane case  there exists an  analogous equation 
\be
\nabla^2_{9-p} y=\tilde\alpha,
\ee
where $C=y(r) \beta$. It is easy to convince one-self that this equation is 
solved for $y=r^2$ for which  $\tilde\alpha$ is constant, and the geometry 
for the $9-p$ dimensional 
space is $ds^2=dr^2+r^2d\Omega^2_{8-p}$ or eq(\ref{IIA}). 
However, if we take the geometry of 
 $9-p$ dimensional space as 
\be
ds^2=\f{dr^2}{f(r)}+r^2 d\Omega^2_{8-p}
\ee 
which is the transverse part of the spacetime for a non-extremal solution 
to Dp branes in IIB supergravity, then for any power law like solution 
 to $y(r)=r^{n}$, do not makes
$\tilde\alpha$  a constant. Hence, in the non-extremal case it is expected that 
it will be more complicated to find the one form $C$. 
But, unfortunately we have not yet solved the complete set of equations 
of motion.

Also, we shall present the non-relativistic supersymmetry preserving  
solutions for Dp branes as well as for M2 brane, for which the 1-form $C$
vanishes but not the function $h$ as defined in eq(\ref{gen_metric}) for D3 
brane case. In general, we may write down such a metric for Dp branes in 
Type IIB as  
\be
ds^2_E=f^{\f{p-7}{8}}\bigg[2dx^+\bigg(dx^-+C+h(S^{8-p})r^{2z}{dx^+}\bigg)+
dx^2_2+\cdots+dx^2_{p}]+f^{\f{p+1}{8}}[dr^2+r^2 d\Omega^2_{8-p}\bigg],
\ee
with appropriate fluxes and matter fields.

The organization of the paper is as follows. In section 2, we shall give a 
prescription for the construction of the non-relativistic solution for a 
bunch of coincident $Dp$ branes in Type IIB theory and 
in the subsections of this section, we shall write down the solutions case by
case and examine its supersymmetry. In section 3, we shall  give the 
prescription for Type IIA and then in the subsections, we shall write down 
the solutions explicitly case by case and examine the supersymmetry preserved
by it. In section 4, we shall construct the non-relativistic solution for a 
bunch of $D3$ and $D5$ branes on the singular Calabi-Yau, that is the conifold.
In section 5, we shall resolve the singularity and find the solution for a 
smooth Calabi-Yau that is for the deformed conifold and in section 6, we shall
write down the solution for M2 brane for which the 1-form  $C$ vanishes 
but not the function $h$ and  then in section 7, 
we shall present the symmetries preserved these solutions and discuss some 
of the drawbacks of such solutions in section 8 and finally conclude in section 9.  Results of the computation of spin connections and the 1-form $C$ in Cartesian coordinates are presented in the appendices.
  
\section{The Prescription}
The prescription to generate non-relativistic supersymmetry preserving 
solution (with $h=0$) in Type IIB 
theories is  by following  two simple step procedure, however such a simple
prescription may not work for  the case where there is  
intersection of branes, for completeness one may have  to check
all the equations of equations of motion explicitly. But before 
going over to those steps, first  one need to rewrite the solutions with 
relativistic symmetries in terms of  the light cone directions, which is 
defined using the time like coordinate and one spatial non-compact direction
along the brane. Let us rewrite  the solution
to N coincident Dp branes of  type IIB, following \cite{imsy},\footnote{We use a different normalization for the fluxes and the
 solutions in this paper are all written in Einstein frame.}
\bea
\label{coincident_rel_solution}
ds^2_E&=&f^{\f{p-7}{8}}[2dx^+dx^-+dx^2_2+\cdots+dx^2_{p}]+f^{\f{p+1}{8}}
[dr^2+r^2 d\Omega^2_{8-p}],\nn
e^{\Phi}&=&f^{\f{3-p}{4}},~~~F_{p+2}=\ell_1(r) dx^+\w dx^-\w dx_2\w \cdots\w
dx_p\w dr,\nn
\ell_1(r)&=& -\f{f'}{f^2},~~~f=1+\f{f_0}{r^{7-p}},~~~{\rm f_0=constant}
\eea

The steps are: 
(1)first step is to replace 
\be\label{shifting}
dx^-\rightarrow dx^-+C,
\ee
where $C$ is a 1-form defined on the directions perpendicular to the brane 
directions, which means the metric along the world volume
directions, up to a conformal factor becomes
\be
ds^2_{p+1}=2dx^+(dx^-+C)+dx^2_2+\cdots+dx^2_{p}
\ee
(2) Second step is to add an extra piece to the $p+2$-form field strength
\be\label{condition_fieldstrength}
F_{p+2}=\ell_1(r) dx^+\w (dx^-+C)\w dx_2\w \cdots\w
dx_p\w dr+\ell_2(r) dx^+\w dx_2\w \cdots\w dx_p\w dC.
\ee

The imposition of the Bianchi identity, $dF_{p+2}=0$, gives  us the 
restriction  that is 
\be
\f{d\ell_2}{dr}+(-1)^{p+1}\ell_1=0,
\ee
using the fact that $\ell_1=-\f{f'}{f^2}$ gives
\be
\ell_2=(-1)^p\times \f{1}{f}.
\ee

For our simple  choice of one kind of coincident branes, the equations of 
motion to the $F_{p+2}$ form field is
\be
d\star_{10}\bigg[ e^{\bigg(\f{3-p}{2}\bigg)\Phi} F_{p+2}\bigg]=0.
\ee
Now, using the relation between the dilaton, $\Phi$ and $f$ as written in 
eq(\ref{coincident_rel_solution}), and solving equations of motion of the 
flux gives the condition on the 1-form $C$
\be\label{harmonic}
d{ \star}_{9-p}dC=0,
\ee
whether the ${ \star}$ is taken with respect to the $9-p$ dimensional
Ricci flat metric
\be\label{perp_metric}
ds^2_{9-p}=dr^2+r^2\Omega^2_{8-p}.
\ee

For $D3$ branes one need to put the extra condition that is the self duality
constraint on eq(\ref{condition_fieldstrength}) with respect to the changed 
metric. From the first step,
 eq(\ref{shifting}), there follows a symmetry transformation under which 
the combination $dx^-+C$ remains invariant
\be
x^-\rightarrow x^--\Lambda,~~~C\rightarrow C+d\Lambda,
\ee
for some $\Lambda$ defined over eq(\ref{perp_metric}).

The dilaton equations of motion do not give anything new and the 
rest of the equations of motion that of the metric components  need to be 
checked explicitly on a case by case basis.


For completeness, The examples that we shall consider are $D1$ branes, 
$D3$ branes and  $NS5/D5$ branes. 

\subsection{The supersymmetry transformation}
The supersymmetry preserving  conditions for Type IIB are given by the 
vanishing condition of dilatino
\be
\delta\lambda=\f{i}{2}\bigg(\p_M\phi+ie^{\phi}\p_M \chi \bigg)\Gamma^M
\epsilon^{\star}-\f{i}{24}\bigg(e^{-\f{\phi}{2}}H_{M_1M_2M_3}+
ie^{\f{\phi}{2}}F_{M_1M_2M_3} \bigg)\Gamma^{M_1M_2M_3}\epsilon
\ee

and that of the gravitino 
\bea
\delta\psi_M&=&\bigg(\p_M+\f{1}{4}\omega^{ab}_M\Gamma_{ab}\bigg)\epsilon-
\f{i}{1920} F_{M_1M_2M_3M_4M_5}\Gamma^{M_1M_2M_3M_4M_5}\Gamma_{M}\epsilon
+\nn&&\f{1}{96}\bigg(e^{-\f{\phi}{2}}H_{M_1M_2M_3}+
ie^{\f{\phi}{2}}F_{M_1M_2M_3} \bigg)\bigg({\Gamma_M}^{M_1M_2M_3}-
9\delta^{M_1}_M\Gamma^{M_2M_3} \bigg)\epsilon^{\star}.
\eea
Using two Majorana-Weyl spinors $\epsilon_L$ and $\epsilon_R$, we can rewrite
\be
\epsilon=\epsilon_L+i\epsilon_R.
\ee
\subsection{D1 branes}

The explicit form of  the solution after following the above 
prescription and checking the equations of motion to metric components
 explicitly, 
for a bunch of coincident $D1$ branes, which are electrically 
charged \footnote{To avoid the 
cluttering of  trigonometric functions,  we shall use a  short hand 
notation: $ sin^2\mu := s^2_{\mu},  ~cos^2\alpha:=c^2_{\alpha}$ etc.}  
\bea\label{d1_sol}
ds^2_E&=&f^{-\f{3}{4}}\bigg[2dx^+(dx^-+C) \bigg]+
f^{\f{1}{4}}\bigg[dr^2+r^2 d\mu^2+ r^2
  s^2_{\mu}d\alpha^2+
  \f{r^2}{4}s^2_{\mu}s^2_{\alpha}(\sigma^2_1+\sigma^2_2+c^2_{\alpha}\sigma^2_3)
+\nn&&\f{r^2}{4} s^2_{\mu}c^2_{\mu}(d\lambda+s^2_{\alpha}\sigma_3)^2+\f{r^2}{4}
 (d\chi+s^2_{\mu}(d\lambda+s^2_{\alpha}\sigma_3))^2 \bigg],\nn
F_3&=& L_1(r) dx^+\w (dx^-+C)\w dr+L_2(r) dx^+\w dC,~~~
C= y(r)\bigg[d\chi+s^2_{\mu}(d\lambda+s^2_{\alpha}\sigma_3)\bigg],\nn
y(r)&=& \sigma r^2,~~~\Phi=\f{1}{2}Log~f(r),~~~f(r)=1+\f{f_0}{r^6},~~~
L_1=-\f{f'}{f^2},~~~L_2=-\f{1}{f},
\eea  

where $f_0$ is a constant, and  $\sigma_i$'s are the SU(2) left 
invariant 1-forms. 
\be\label{sigma_1_forms}
\sigma_1=c_{\psi}d\theta+s_{\psi}s_{\theta}d\phi,~
\sigma_2=-s_{\psi}d\theta+c_{\psi}s_{\theta}d\phi,~
\sigma_3=d\psi+c_{\theta}d\phi.
\ee

Upon computing the dilatino and gravitino variation using the spin-connections as  written in eq(\ref{spin_connections_d1brane}), with the choice 
\be
\epsilon^{\star}=-i\epsilon,
\ee
requires the following condition on the spinor
\be
\Gamma^+\epsilon=0
\ee 

From which it follows that  the above solution of $D1$ brane breaks
one half of the supersymmetry.  The form of the spinor 
\be
\epsilon=e^{\f{9}{8}Log~r}\times \epsilon(R^8),
\ee
where the spinor $\epsilon(R^8)$ is defined over the flat $R^8$, and the $S^7$
part of it is described as a U(1) fibration over $CP^3$. 

The expression to, 1-form $C$, is presented in Cartesian coordinates in eq(\ref{C_complex_coordinate_s7}) and eq(\ref{C_complex_coordinate}).

\subsubsection{Solution with $h\neq 0$}

The D1 brane solution with $h\neq 0$ but with $C=0$
\bea
ds^2_E&=&f^{-\f{3}{4}}\bigg[2dx^+dx^- +2h(S^7) r^{2z}{dx^+}^2\bigg]+
f^{\f{1}{4}}\bigg[dr^2+r^2 d\mu^2+ r^2
  s^2_{\mu}d\alpha^2+\nn&&
  \f{r^2}{4}s^2_{\mu}s^2_{\alpha}(\sigma^2_1+\sigma^2_2+c^2_{\alpha}\sigma^2_3)
+\f{r^2}{4} s^2_{\mu}c^2_{\mu}(d\lambda+s^2_{\alpha}\sigma_3)^2+\f{r^2}{4}
 (d\chi+s^2_{\mu}(d\lambda+s^2_{\alpha}\sigma_3))^2 \bigg],\nn
F_3&=&-\f{f'}{f^{2}} dx^+\w dx^-\w dr,~~~ 
~~~\Phi=\f{1}{2}Log~f(r),~~~f(r)=1+\f{f_0}{r^6},\nn
-\nabla^2_{S^7}h&=&4z(z+3)h,
\eea
where the function $h(S^7)$ is defined on $S^7$ and the solution 
preserves $1/2$ of the supersymmetry  with the condition on the spinor as
 $\Gamma^+\epsilon=0$

\subsection{D3 branes}

In this case the  solution is given in \cite{mmt} and \cite{dg} for all 
Calabi-Yau's of the singular type, $ds^2(CY_3)=dr^2+r^2ds^2(SE_5)$, but for 
completeness we shall present the result for a specific five dimensional 
$SE_5$ manifold that is for $S^5$ and with $z=4$. The exponent $z=2+\lambda$,
and $\lambda$ is taken from the form of $C=\sigma r^{\lambda}\beta$.

The solution is
\bea\label{d3_sol}
ds^2_E&=&f^{-\f{1}{2}}[2dx^+(dx^-+C)+dx^2_2+dx^2_3]+f^{\f{1}{2}}[dr^2+r^2d\mu^2
+
\f{r^2}{4}s^2_{\mu}(d\theta^2+s^2_{\theta}d\phi^2+c^2_{\mu}\sigma^2_3)
+\nn&&\f{r^2}{4}(d\chi+s^2_{\mu}\sigma_3)^2],~~~
F_5=(1+\star_{10}){\cal F}_5,~~~C= y(r)[d\chi+s^2_{\mu}\sigma_3]\nn
{\cal F}_5&=&L_1(r)dx^+\w (dx^-+C)\w dx_2\w
  dx_3\w dr+L_2(r) dx^+\w dx_2\w dx_3\w dC,\nn
f(r)&=&1+\f{f_0}{r^4},~~~L_1=-\f{f'}{f^2},~~~L_2=-\f{1}{f},~~~y(r)=\sigma r^2.
\eea
We shall check the  supersymmetry preserved of this solution only in the near 
horizon 
limit where we shall drop ``1'' from the harmonic function $f$, for simplicity 
of the calculation.

The computation of the dilatino variation, using the spin-connections eq(\ref{spin_connections_d3brane}), do not give any information and the
gravitino variation gives
\be\label{projection_condition_d3}
\Gamma^+\epsilon=0,~~~\Gamma^{+-x_2x_3}=-i\epsilon.
\ee

Hence the non-relativistic $D3$ brane  solution breaks 
one quarter of  supersymmetries. The $\Gamma^{11}$ is defined 
\be
\Gamma^{11}=\Gamma^{+-x_2x_3r\mu\theta\phi\sigma_3\chi},~~~
\Gamma^{11}\epsilon=\epsilon.
\ee
the form of the Killing spinor is 
\be
\epsilon=e^{-\f{1}{8}Log~f}\times\epsilon(R^6),
\ee
where the spinor $\epsilon(R^6)$ is defined over the flat $R^6$, but  
written in 
spherical polar coordinate system. The $S^5$ of it is written as U(1) 
fibration over the $CP^2$. 

It may be noted that the negative sign that we get in the right hand side of
eq(\ref{projection_condition_d3}) is due to the choice of signs that appeared
in the five-form flux.  The expression to, 1-form $C$, is presented in Cartesian coordinates in eq(\ref{C_complex_coordinate_s5}) and eq(\ref{C_complex_coordinate}).

\subsubsection{Solution with $h\neq 0$} 

The D3 brane solution is already presented in \cite{hy}, let us present it,
 for completeness
\bea
ds^2_E&=&f^{-\f{1}{2}}\bigg[2dx^+dx^-+2h(S^5)r^{2z-2}{dx^+}^2+dx^2_2+dx^2_3
\bigg]+
f^{\f{1}{2}}\bigg[dr^2+r^2d\mu^2+\nn&&
\f{r^2}{4}s^2_{\mu}(d\theta^2+s^2_{\theta}d\phi^2+c^2_{\mu}\sigma^2_3)
+\f{r^2}{4}(d\chi+s^2_{\mu}\sigma_3)^2\bigg],~~~
F_5=(1+\star_{10}){\cal F}_5,\nn
{\cal F}_5&=&-\f{f'}{f^2}dx^+\w dx^-\w dx_2\w
  dx_3\w dr,~~~f(r)=1+\f{f_0}{r^4},\nn
\nabla^2_{S^5}h&=&-4(z^2-1)h,
\eea
where the function $h(S^5)$ is defined on  $S^5$ and the solution preserves
 $1/4$ of supersymmetry with the conditions on spinor as
\be
\Gamma^+\epsilon=0,~~~\Gamma^{+-x_2x_3}=-i\epsilon.
\ee

\subsection{NS5 branes}

The form of the magnetically charged solution, in this case in Einstein 
frame is
\bea\label{ns5_sol}
ds^2_E&=&f^{-\f{1}{4}}\bigg[ 2dx^+(dx^-+C)+dx^2_2+dx^2_3+dx^2_4+dx^2_5\bigg]+
f^{\f{3}{4}}[dr^2+\nn &&
\f{r^2}{4}(d\theta^2+d\phi^2+d\psi^2+2c_{\theta}d\phi
  d\psi)],~~~C= y(r)[d\psi+c_{\theta}d\phi],~y(r)=\sigma r^2\nn 
H_3&=& \f{f_0}{4}s_{\theta}d\theta\w d\phi\w d\psi+dx^+\w dC,~~~
\Phi=\f{1}{2}Log~f,~~~f(r)=1+\f{f_0}{r^2}.
\eea

The magnetically charged solution to $D5$ branes can be generated by 
S-dualising the above solution and reads as
\bea
ds^2_E&=&f^{-\f{1}{4}}\bigg[ 2dx^+(dx^-+C)+dx^2_2+dx^2_3+dx^2_4+dx^2_5\bigg]+
f^{\f{3}{4}}[dr^2+\nn &&
\f{r^2}{4}(d\theta^2+d\phi^2+d\psi^2+2c_{\theta}d\phi
  d\psi)],~~~C= y(r)[d\psi+c_{\theta}d\phi],~y(r)=\sigma r^2\nn 
F_3&=& -\f{f_0}{4}s_{\theta}d\theta\w d\phi\w d\psi-dx^+\w dC,~
\Phi=-\f{1}{2}Log~f,~~~f(r)=1+\f{f_0}{r^2}.
\eea 
In order to check the supersymmetry preserved by this solution, which 
we shall do only in the near horizon limit.   Let us take a choice
\be
\epsilon^{\star}=-i\epsilon.
\ee
For which the supersymmetric conditions for 
the near horizon solution to $D5$ brane  may suggests the following 
condition on the spinors 
\be
\Gamma^+\epsilon=0,~~~ 
\ee
where the $\Gamma^+$ matrix is defined on the tangent space and $\epsilon$
is the  spinor for the relativistic theory. Let us check that by doing 
explicit computation.

 

Let us check explicitly the supersymmetry preserved by the above solution. The 
dilatino variation equation using eq(\ref{spin_connections_ns5brane}), gives the condition 
\be
\Gamma^+\epsilon=0,~~~\bigg(\Gamma^r-\Gamma^{\theta\phi\psi}\bigg)\epsilon=0.
\ee
The gravitino variation gives
\be
\Gamma^+\epsilon=0,~
\p_r\epsilon-\f{1}{8fr^3}\epsilon=0,~
\epsilon\neq \epsilon(x^+,~x^-,~x_i,\theta,\phi)
\ee
and the form of the spinor is
\be
\epsilon=e^{\f{log~r}{8}} \times
e^{-\f{\psi}{2}\Gamma^{\theta\phi}}\epsilon^0
\ee
where $\epsilon^0$ is a constant spinor.
From this it just follows that the non-relativistic $D5$ brane breaks one 
quarter of the   supersymmetry. The expression to, 1-form $C$, is presented in Cartesian coordinates in eq(\ref{C_complex_coordinate_s3}) and eq(\ref{C_complex_coordinate}).

\subsubsection{Solution with $h\neq 0$}

The D5 brane solution 

\bea\label{d5_sol}
ds^2_E&=&f^{-\f{1}{4}}\bigg[ 2dx^+dx^-+2h(S^3)r^{2z}{dx^+}^2+dx^2_2+dx^2_3+
dx^2_4+dx^2_5\bigg]+f^{\f{3}{4}}\bigg[dr^2+\nn&&
\f{r^2}{4}(d\theta^2+d\phi^2+d\psi^2+2c_{\theta}d\phi
  d\psi)\bigg],\nn 
F_3&=& -\f{f_0}{4}s_{\theta}d\theta\w d\phi\w d\psi,~
\Phi=-\f{1}{2}Log~f,~~~f(r)=1+\f{f_0}{r^2},\nn
\nabla^2_{S^3}h&=&-4z(z+1)h,
\eea 
where the function $h(S^3)$ is defined on $S^3$ and the solution preserves 
$1/4$ of supersymmetries and the conditions on spinor are
\be
\Gamma^+\epsilon=0,~~~\Gamma^{r\theta}\epsilon=-\Gamma^{\phi\psi}\epsilon.
\ee

\section{Type IIA solutions}
In this section we shall try to generate solutions with even value of 
$p$ to Dp branes 
i.e. in Type IIA theory. As before, if we want the 1-form $C$ to still obey 
 eq(\ref{harmonic_condition}) then we need to deform the $9-p$ dimensional 
space transverse to the brane
directions even though it is still Ricci flat. The prescription is that, we
shall take the $9-p$ dimensional  space to be the direct product of a $8-p$ 
dimensional space and $S^1$. For this particular choice, we can very easily 
define the 1-form $C$. As even dimensional spheres cannot be written as 
 $U(1)$ fibration over the complex projective spaces.

Now, in order to have a non-relativistic solution with the condition 
eq(\ref{harmonic_condition}), let us proceed as follows.  
Let us take the ansatz to the solution as before but with a simple 
modification to the topology of metric transverse to the brane direction as
$R^1\times S^1\times S^{7-p}$, for even $p$ 
\bea
ds^2_E&=&f^{\f{p-7}{8}}[2dx^+(dx^-+C)+dx^2_2+\cdots+dx^2_{p}]+f^{\f{p+1}{8}}
[dr^2+r^2 d\Omega^2_{7-p}+dw^2],\nn
e^{\Phi}&=&f^{\f{3-p}{4}},~~~
F_{p+2}=\ell_1(r) dx^+\w (dx^-+C)\w dx_2\w \cdots\w dx_p\w dr+\nn 
&&\ell_2(r) dx^+\w dx_2\w\cdots\w dx_p\w dC 
\eea
where the $w$ is the  coordinate  used to describe  $S^1$. 
Imposing the Bianchi identity on $F_{p+2}$ gives
\be
\f{d\ell_2(r)}{dr}+(-1)^{p+1}\ell_1(r)=0
\ee
and as before the equation of motion of $F_{p+2}$ form field strength gives
eq(\ref{harmonic_condition}). The only other difference this time is the 
structure of the harmonic function $f(r)$

\be
f(r)=1+\f{f_0}{r^{6-p}},
\ee
where $f_0$ is a constant.
\subsection{Supersymmetric variations }
The supersymmetric conditions for Type IIA  theory  are determined by the 
vanishing of  the dilatino $\lambda$ and gravitino $\psi_M$ variations   and 
are
\bea
\delta\lambda&=&\f{1}{2}\p_M \phi\Gamma^M \Gamma^{11}\epsilon+\f{3}{16}
e^{3/4 \phi} F_{M_1M_2}\Gamma^{M_1M_2}\epsilon+\f{i}{24} e^{-\f{\phi}{2}}
H_{M_1M_2M_3}\Gamma^{M_1M_2M_3}\epsilon-\nn&&\f{i}{192}e^{\f{\phi}{4}}
F_{M_1M_2M_3M_4}\Gamma^{M_1M_2M_3M_4}\epsilon,
\eea
\bea
\delta\psi_M&=&(\p_M+\f{1}{4}\omega^{ab}_M\Gamma_{ab})\epsilon+\f{1}{64}
e^{\f{3\phi}{4}}F_{M_1M_2}\bigg({\Gamma_M}^{M_1M_2}-
14\delta^{M_1}_M\Gamma^{M_2}\bigg)\Gamma^{11}\epsilon+\nn&&\f{1}{96}
e^{-\f{\phi}{2}}H_{M_1M_2M_3}\bigg({\Gamma_M}^{M_1M_2M_3}-
9\delta^{M_1}_M\Gamma^{M_2M_3}\bigg)\Gamma^{11}\epsilon+\nn&&
\f{i}{256}
e^{\f{\phi}{4}}F_{M_1M_2M_3M_4}\bigg({\Gamma_M}^{M_1M_2M_3M_4}-
\f{20}{3}\delta^{M_1}_M\Gamma^{M_2M_3M_4}\bigg)\Gamma^{11}\epsilon
\eea

\subsection{D2 branes}
In this subsection, we shall give  the   electrically charged 
solution to $D2$ branes
and write down the explicit structure to  1-form $C$ and it reads  
in Einstein frame  
\bea
ds^2_E&=&f^{-\f{5}{8}}[2dx^+(dx^-+C)+dx^2_2]+
f^{\f{3}{8}}[dr^2+r^2d\mu^2
+\f{r^2}{4}s^2_{\mu}(d\theta^2+s^2_{\theta}d\phi^2+c^2_{\mu}\sigma^2_3)
+\nn&&\f{r^2}{4}(d\chi+s^2_{\mu}\sigma_3)^2+dw^2],\nn
F_4&=& L_1(r) dx^+\w(dx^-+C)\w dx_2\w dr+L_2(r) dx^+\w dx_2\w dC,\nn
C&=& y(r) [d\chi+s^2_{\mu}\sigma_3], ~~~\Phi=\f{1}{4}Log~f,~~~
f(r)=1+\f{f_0}{r^4},\nn
L_1&=&-\f{f'}{f^2},~~~L_2=\f{1}{f(r)},~~~y(r)=\sigma r^2
\eea

From the variation of  gravitino and dilatino using eq(\ref{spin_connections_d2brane}), we get the following conditions on the spinor
\be
\Gamma^+\epsilon=0,~~~\Gamma^{+-x_2}\epsilon=-i\epsilon,
\ee 
and the $\Gamma^{11}$ matrix is defined as 
\be
\Gamma^{11}=\Gamma^{+-x_2r\mu\theta\phi\psi\chi w},~~~
\Gamma^{11}\epsilon=\epsilon.
\ee

From the supersymmetric variation to gravitino and dilatino, it follows that the
non-relativistic $D2$ brane solution breaks one quarter of the supersymmetry
and the form of the spinor, in the near horizon limit 
\be
\epsilon=e^{\f{5}{8}Log~r}\epsilon(R^6),
\ee
where $\epsilon(R^6)$ is defined over the flat $R^6$ but written in spherical 
polar coordinate system, where  the $S^5$ is written as a U(1) fibration over 
the complex projective space $CP^2$.

It is interesting to see that if take the coordinate $w$ as a non-compact 
coordinate then the geometry to D2 branes in the string frame
do obeys the scaling symmetries as written in eq(\ref{scaling_symmetry}) 
along with $w\rightarrow \f{w}{\mu}$, however the four-form flux, 
$F_4$, breaks it.  

\subsubsection{Solution with $h \neq 0$}

The D2 brane solution reads
\bea
ds^2_E&=&f^{-\f{5}{8}}[2dx^+dx^-+2h(S^5)r^{2z}{dx^+}^2+dx^2_2]+
f^{\f{3}{8}}[dr^2+r^2d\mu^2+\nn&&
\f{r^2}{4}s^2_{\mu}(d\theta^2+s^2_{\theta}d\phi^2+c^2_{\mu}\sigma^2_3)+
\f{r^2}{4}(d\chi+s^2_{\mu}\sigma_3)^2+dw^2],\nn
F_4&=&-\f{f'}{f^2} dx^+\w dx^-\w dx_2\w dr,
 ~~~\Phi=\f{1}{4}Log~f,~~~
f(r)=1+\f{f_0}{r^4},\nn
\nabla^2_{S^5}h&=&-4z(z+2)h,
\eea
where the function $h(S^5)$ is defined on $S^5$ and the above solution 
preserves $1/4$ of the supersymmetry and the conditions on spinors are
\be
\Gamma^+\epsilon=0,~~~\Gamma^{+-x_2}\epsilon=-i\epsilon.
\ee

\subsection{D4 branes}
The magnetically charged  solution to D4 branes with the explicit structure 
to  1-form $C$ is
\bea
ds^2_E&=&f^{-\f{3}{8}}[2dx^+(dx^-+C)+dx^2_2+dx^2_3+dx^2_4]+\nn &&
f^{\f{5}{8}}[dr^2+\f{r^2}{4}(d\theta^2+d\phi^2+d\psi^2+2c_{\theta}d\phi
  d\psi)+dw^2],\nn
F_4&=&\f{f_0}{4}s_{\theta}d\theta\w d\phi\w d\psi\w dw+dx^+\w dC\w dw,~~~
C= y(r)[d\psi+c_{\theta}d\phi],\nn
\Phi&=&-\f{1}{4}Log~f(r),~~~f(r)=1+\f{f_0}{r^2},~~~y(r)=\sigma r^2
\eea


While checking the supersymmetry preserved by the solution, we shall use 
the near horizon solution. 
The dilatino and gravitino variation using eq(\ref{spin_connections_d4brane}) gives the conditions on spinor
\be
\Gamma^+\epsilon=0,~~~\Gamma^{+-x_2x_3x_4}\epsilon=-i\epsilon,
\ee

Which suggests that the solution breaks $1/4$ of the supersymmetry.\\

In doing the calculation we have defined  $\Gamma^{11}$  as 
\be
\Gamma^{11}=\Gamma^{+-x_2x_3x_4r\theta\phi\psi z},
\ee
where again the $\Gamma^M$ matrices are defined in the tangent space and the 
form of the spinor is
\be
\epsilon
=e^{\f{3}{16}Logr}\times \epsilon(R^4),
\ee
where $\epsilon^0$ is a constant spinor and $\epsilon(R^4)$ is the spinor 
defined on flat $R^4$, which depends on the angles of the $S^3$. The metric
on $S^3$ is written as a U(1) fibration over $CP^1$.

\subsubsection{Solution with $h \neq 0$}

The D4 brane solution 

\bea
ds^2_E&=&f^{-\f{3}{8}}[2dx^+dx^-+2h(S^3)r^{2z}{dx^+}^2+dx^2_2+dx^2_3+dx^2_4]+
f^{\f{5}{8}}[dr^2+\nn &&\f{r^2}{4}(d\theta^2+d\phi^2+d\psi^2+2c_{\theta}d\phi
  d\psi)+dw^2],\nn
F_4&=&\f{f_0}{4}s_{\theta}d\theta\w d\phi\w d\psi\w dw,~~~
\Phi=-\f{1}{4}Log~f(r),~~~f(r)=1+\f{f_0}{r^2},\nn
\nabla^2_{S^3}h&=&-4z(z+1)h,
\eea
where the function $h(S^3)$ is defined on $S^3$ and the solution preserves 
$1/4$ of supersymmetry and the conditions on spinors are
\be
\Gamma^+\epsilon=0,~~~\Gamma^{+-x_2x_3x_4}\epsilon=-i\epsilon.
\ee

\section{D3 and D5 branes on the singular Conifold}
In this section, we shall find the solution of D3 and D5 branes put at the 
singularity of the Calabi-Yau that is the conifold, where the D5 brane is 
wrapped over the $S^2$ of it. 
In order to find supersymmetry preserving solution in 10 dimensional 
supergravity with  a Calabi-Yau,  
the prescription of \cite{dg} is to find a one form object, $C$, such that 
it obeys the following equation
\be
d\star_{CY}dC=0,~(d^{\dagger}d+dd^{\dagger})_{SE}\beta=\mu\beta,~d^{\dagger}\beta=0
\ee
where
\be
C=y(r)\beta,~~y(r)=\f{\sigma}{3} r^{\lambda},~~\mu=\lambda(\lambda+2),
\ee
where $\sigma$ is a constant and  the metric on the Calabi-Yau is that of
 a cone over a 5 dimensional Sasaki-Einstein base space
\be
ds^2(CY_3)=dr^2+r^2ds^2(SE_5)
\ee

Let us take a specific choice to $\lambda=2$, which says $\mu=8$. For this 
choice of $\lambda$, let us take the Calabi-Yau as the conifold and with the 
base space is that of $T^{1,1}$. In this case the metric of the $T^{1,1}$ 
is described as \cite{cd}
\be
ds^2(T^{1,1})=\f{1}{6}(g^2_1+g^2_2+g^2_3+g^2_4)+\f{g^2_5}{9},
\ee
where the one-forms $g_i$'s are defined as 
\be
g_1=\f{e_1-e_3}{\sqrt{2}},~g_2=\f{e_2-e_4}{\sqrt{2}},~
g_3=\f{e_1+e_3}{\sqrt{2}},~g_4=\f{e_2+e_4}{\sqrt{2}},~g_2=e_5
\ee
with
\bea
e_1&=&-sin\theta_1d\phi_1,~~e_2=d\theta_1,~~e_3=cos\psi sin\theta_2 d\phi_2-
sin\psi d\theta_2,\nn 
e_4&=&sin\psi sin\theta_2 d\phi_2+cos\psi d\theta_2,~~
e_5=d\psi+cos\theta_1 d\phi_1+cos\theta_2 d\phi_2.
\eea

Using the condition that $dC$ is co-closed on the conifold for $\lambda=2$ 
gives 
\be
\beta=g_5=d\psi+cos\theta_1 d\phi_1+cos\theta_2 d\phi_2
\ee

Let us try to construct a solution in Type IIB on the conifold with $N D3$ 
branes as well as $M D5$ branes where the latter kind of branes are being 
wrapped on the $S^2$ of the $T^{1,1}$ with the rest of the directions of $D5$ 
and $D3$ branes are extended along $x^+,~x^-,~x_2,~x_3$ directions. 
 
The ansatz that we shall take is 
\bea
ds^2&=&h^{-\f{1}{2}}[2dx^+(dx^-+C)+dx^2_2+dx^2_3]+h^{\f{1}{2}}
[dr^2+\f{r^2}{6}(g^2_1+g^2_2+g^2_3+g^2_4)+\f{r^2}{9}g^2_5],\nn
F_5&=&(1+\star_{10}){\cal F}_5,~~~\Phi=Log[g_s],~~~C_0=0,\nn
{\cal F}_5&=&\ell_1(r)h^{-\f{3}{4}}
dx^+\w (dx^-+C)\w dx_2\w dx_3\w dr-\ell_2(r) h^{-\f{3}{4}}dx^+\w dx_2\w dx_3\w 
dC,\nn
F_3&=&\f{M\alpha'}{4}g_5\w(g_1\w g_2+g_3\w g_4),~~~
H_3=\f{g_sM\alpha'}{2}f' dr\w (g_1\w g_2+g_3\w g_4),
\eea
and substituting it into the equations of motion of, $\Phi$, dilaton gives us 
the following solution to $f(r)$
\be
f(r)=\f{3}{2}Log[r/r_0]
\ee 
It is easy to check that the ISD conditions on the complex combination of the 
3-form fluxes is  still there. 

The equations of motion of $F_3$-form flux gets identically satisfied, whereas 
the $H_3$-form flux gives
\be
\f{d}{dr}\bigg(\f{rf'}{h} \bigg)=\f{3}{2}g_s\ell_1 h^{-\f{3}{4}}.
\ee
The 5-form flux, $F_5$, gives us the following equations
\be
\f{g_sM^2\alpha^{'2}}{4}f'=\f{1}{108}\f{d}{dr}
\bigg(\ell_1 r^5 h^{\f{5}{4}} \bigg),~~~
\ell_1 h^{-\f{3}{4}}=\f{d}{dr}
\bigg(\ell_2  h^{-\f{3}{4}} \bigg),~~~
4\ell_2 r^3 h^{\f{1}{4}}=\f{d}{dr}
\bigg(\ell_2 r^4 h^{\f{1}{4}} \bigg),  
\ee

Using the expression of $f(r)$, results in 
\be
\ell_1(r)=-\f{h'}{g_sh^{\f{5}{4}}},~~~\ell_2(r)=\f{1}{g_s h^{\f{1}{4}}}
\ee
and 
\be
h(r)=\f{27\pi\alpha^{'2}}{4r^4}\bigg[g_sN+\f{3}{2\pi}(g_sM)^2 Log~r+
\f{3}{8\pi}(g_sM)^2\bigg]
\ee

The equations of motion to the metric components are all satisfied using 
the following relation that the warp factor satisfies
\be
5h'+rh''=-\f{81}{2}\f{g^2_sM^2\alpha^{'2}}{r^5}.
\ee
The convention that we have adopted in finding the Hodge duals is 
$\epsilon_{+-x_2x_3r12345}=-1$.

It is interesting to note that for $M=0$, we reproduced the solution 
written in \cite{dg} for the
conifold and for $\sigma=0$, we get back the Klebanov-Tseytlin solution 
\cite{kt}. Moreover,  
it is interesting to note that $\sigma$, neither appeared in the 3-form fluxes
nor in the warp factor. So, the  deformation of the relativistic solution 
to generate non-relativistic solution is done in such a way that the 
three form fluxes, dilaton, warp factor of the  final 
solution do not depends on the deformation parameter $\sigma$.

We expect  that this is true for any value of $\lambda$ with the
appropriate one-form $\beta$
which is the eigenfunction  of the Laplacian as  defined above
and  still 
solves the equations of motion of Type IIB,  with the same choice of the 
3-form field strengths as written in the ansatz.

\section{D3 and D5 branes on the deformed conifold}

In this section, we shall find the non-relativistic solution to
Klebanov-Strassler configuration \cite{ks} and 
 construct the one form  $C$, which is required to  satisfy the following 
condition on the Calabi-Yau
\be
\label{harmonic}
d\star_{CY_3}dC=0
\ee
 where the metric of the deformed conifold is 
\be
ds^2_6=\f{1}{2}\varepsilon^{\f{4}{3}}K(\tau)
\bigg[\f{d\tau^2+g^2_5}{3K(\tau)^3}+
cosh^2~(\tau/2)(g^2_3+g^2_4)+sinh^2(\tau/2)(g^2_1+g^2_2) \bigg]
\ee
where with a slight change of notation we are denoting the radial coordinate 
as $\tau$
in stead of $r$, and  will represent one of the direction perpendicular to 
the brane direction  and  the function 
\be
K(\tau)=\f{(sinh2\tau-2\tau)^{\f{1}{3}}}{2^{\f{1}{3}}sinh\tau}
\ee

Let us assume that the object $C$ takes the following form: $C=y(\tau)g_5$,
 and $y(\tau)$ is going to be determined by 
eq(\ref{harmonic}). The solution to\footnote{A similar equation is  
solved in \cite{ghk}, while the authors were 
investigating the presence of axionic strings in cascading gauge theories.} 
\be
y(\tau)=\sigma (sinh(2\tau)-2\tau)^{\f{1}{3}},~~~{\rm Or}~
y(\tau)=\sigma (sinh(2\tau)-2\tau)^{-\f{2}{3}},
\ee
where $\sigma$ is a constant integration. In the following we shall use 
a notation
\be
\alpha: =\f{\varepsilon^{\f{2}{3}}}{\sqrt{2}},~~~
\beta: =\f{\varepsilon^{\f{2}{3}}}{\sqrt{6}}
\ee

Note that this $\beta$ should not be confused with the 1-form $\beta$ that is 
defined in the introductory section. It is interesting to find that the
$\tilde\alpha$ that is defined in eq(\ref{phi_alpha}), vanishes only for the 
second $y(\tau)$.

Let us consider a configuration of $N D3$ branes and $M D5$ branes, where 
the latter  are wrapped around the $S^2$ of the deformed conifold and 
extended along the non-compact directions.  Now, let us  take the following 
ansatz to the solution 
\bea
ds^2&=&h^{-\f{1}{2}}(\tau)[2dx^+(dx^-+C)+dx^2_2+dx^2_3]+
h^{\f{1}{2}}(\tau)ds^2_6,~~~
\Phi=Log[g_s],~~~C_0=0,\nn
F_5&=&(1+\star_{10})\f{\ell_1(\tau)\beta h^{-\f{3}{4}}}{K}
[dx^+\w(dx^-+C)\w dx_2\w dx_3\w d\tau-\nn&&
y'\ell_2(\tau)h^{-\f{3}{4}}dx^+\w dx_2\w
dx_3\w d\tau \w g_5+y\ell_2(\tau)h^{-\f{3}{4}}dx^+\w dx_2\w dx_3\w
(g_1\w g_4-g_2\w g_3) ],\nn 
F_3&=&\f{M\alpha'}{2}[(1-F)g_5\w g_3\w g_4+F g_5\w g_1\w g_2+
F'd\tau\w(g_1\w g_3+g_2\w g_4)],\nn 
H_3&=&\f{g_sM\alpha'}{2}[d\tau\w(f'g_1\w g_2+k'g_3\w g_4)+\f{k-f}{2}g_5\w
(g_1\w g_3+g_2\w g_4)],
\eea

where $'$ denotes derivative with respect to $\tau$. With this choice to 
ansatz 
the equations that follow from the 5-form flux are
\bea\label{5-form_eom}
&&\f{d}{d\tau}[\ell_2 h^{-\f{3}{4}} ]=\f{\ell_1\beta h^{-\f{3}{4}}}{K},~~~
\ell_1 K\alpha^4\beta h^{\f{5}{4}}sinh^2\tau=g_sM^2\alpha^{'2}\ell,\nn
&&\f{d}{d\tau}[y'\alpha^4K^4\ell_2 h^{\f{1}{4}}sinh^2\tau]=8\beta^4y\ell_2 
h^{\f{1}{4}},~~~\ell:=f(1-F)+kF
\eea
The, $\Phi$, dilaton equation gives
\be
\f{f'^2}{sinh^4(\tau/2)}+\f{k'^2}{cosh^4(\tau/2)}+2\f{(k-f)^2}{sinh^2\tau}=
\f{(1-F)^2}{cosh^4(\tau/2)}+\f{F^2}{sinh^4(\tau/2)}+8\f{F'^2}{sinh^2\tau}
\ee
The $F_3$ form field gives
\be
(1-F)tanh^2(\tau/2)-F coth^2(\tau/2)+2h \f{d}{d\tau}\bigg(\f{F'}{h}\bigg)=
\f{g_s(k-f)h}{K}\ell_1\beta h^{-\f{3}{4}}.
\ee

The $H_3$, form field gives
\bea
&&\f{d}{d\tau}\bigg(\f{f'}{h}coth^2(\tau/2) \bigg)+\f{k-f}{2h}=
g_s (1-F)\f{\ell_1\beta}{K} h^{-\f{3}{4}},\nn
&&\f{d}{d\tau}\bigg(\f{k'}{h}tanh^2(\tau/2) \bigg)-\f{k-f}{2h}=
g_s F\f{\ell_1\beta}{K} h^{-\f{3}{4}}
\eea

The equation of motion associated to axion, $C_0$,  is identically satisfied.
Upon solving these equations results in 
\be
\ell_1=-\f{h' K}{\beta h^{\f{5}{4}}},~~~\ell_2=\f{1}{g_sh^{\f{1}{4}}},~~~
h'=-\f{(g_sM\alpha')^2\ell}{K^2\alpha^4sinh^2\tau}
\ee
and
\bea
F&=&\f{sinh\tau-\tau}{2sinh\tau},~~~f=\f{\tau coth\tau-1}{2sinh\tau}
(cosh\tau-1),~~~k=\f{\tau coth\tau-1}{2sinh\tau}(cosh\tau+1),\nn
\ell&=&\f{\tau coth\tau-1}{4sinh^2\tau}(sinh2\tau-2\tau)
\eea

The  condition eq(\ref{harmonic}) is used in satisfying the $F_5$ form 
equations of motion, especially the last equation of (\ref{5-form_eom}).
It just follows that the one-form $C$  is independent of 
 whether we have got  trivial  or non-zero 3-from fluxes.
 
It is also interesting to note that the ISD condition on the complex 
combination of the 3-form fluxes is
 still satisfied and it suggests that the above solution is 
supersymmetric because 
with respect to the complex structure of the deformed conifold the complex
combination of the 3-form flux is of the type (2,1).

The explicit computations of the equations of motion to metric  components can 
identically be satisfied. In fact the 
computation of the Ricci scalar from the ansatz to metric 
says that it is independent of $\sigma$ and which should be the case.

\section{M2 brane solution}

The non-relativistic M2 brane solution with vanishing 1-form $C$
\bea
ds^2&=&f^{-\f{2}{3}}\bigg[2dx^+dx^-+2h(S^7)r^{2z-2}{dx^+}^2+dx^2_2\bigg]+
f^{\f{1}{3}}\bigg[dr^2+r^2 d\mu^2+ r^2 s^2_{\mu}d\alpha^2+\nn &&
  \f{r^2}{4}s^2_{\mu}s^2_{\alpha}(\sigma^2_1+\sigma^2_2+c^2_{\alpha}\sigma^2_3)
+\f{r^2}{4} s^2_{\mu}c^2_{\mu}(d\lambda+s^2_{\alpha}\sigma_3)^2+\f{r^2}{4}
 (d\chi+s^2_{\mu}(d\lambda+s^2_{\alpha}\sigma_3))^2 \bigg],\nn
F_4&=&-\f{f'}{f^{2}}dx^+\w dx^-\w dx_2\w dr,~~~f(r)=1+\f{f_0}{r^6},\nn
\nabla^2_{S^7}h &=&-4(z-1)(z+2)h,
\eea
where $h(S^7)$ is defined on $S^7$ and  preserves $1/4$ of the supersymmetry
and the conditions on spinor are
\be
\Gamma^+\epsilon=0,~~~\Gamma^{+-x_2}\epsilon=-\epsilon.
\ee

In the near horizon limit, where we drop ``1'' in the harmonic function, 
the solution shows a scaling symmetry and the explicit structure of it looks as
\be
r\rightarrow \f{r}{\mu},~x^+\rightarrow \mu^{1+z}x^+,~
x^-\rightarrow \mu^{3-z} x^-,~x_2\rightarrow \mu^2 x_2.
\ee

\section{Symmetries}

The continuous symmetries  of the above solutions includes the symmetries of
the Galilean algebra.

The generators are that of time translation, $H$, spatial translations, 
$P_i$, the Galilean boosts, $K_i$, the rotations in the $x_i$ plane, 
$M_{ij}$,  and the rest mass operator, $N$. 
The explicit structure to the Galilean boost symmetry transformations is
\be
x_i\rightarrow x_i-v_i x^+,~~~x^-\rightarrow x^-+v_i x_i-\f{1}{2} v_iv_i x^+
\ee   

In fact, this particular structure to  Galilean boosts matches with that of in
\cite{bm}.\\

The Dp branes for $p\neq 3$ break explicitly the scale invariance as well as 
the special conformal transformations, which means the 
solutions presented above may be interpreted as the non-conformal Galilean
branes. However,  for $D3$ branes with $h=0$ and $C\neq 0$,
the  dynamical exponent $z$ equals to $4$
and  the scaling symmetries acts as \cite{dg}
\be\label{scaling_symmetry}
r\rightarrow \f{r}{\mu},~~~x^+\rightarrow \mu^4 x^+,~~~ 
x_i\rightarrow \mu x_i, 
~~~(x^-,~C)\rightarrow \f{1}{\mu^2}(x^-,~C),
\ee  
and the discrete symmetries in this case are that of only 
$x_i\rightarrow -x_i$ but there is no 
time reversal symmetry because of the intrinsic Galilean symmetric structure.

For D3 brane case, with $C=0$ and $h\neq 0$, there exists a special conformal 
transformation for $z=2$ whose structure is same as written in 
\cite{bm} but with $r\rightarrow \f{1}{r}$. Also in this case there exists a
scaling symmetry \cite{son},\cite{bm} and \cite{hy}.

\section{Drawback}

Even though the geometries as written above still preserves a fraction of the 
supersymmetry, but unfortunately the mixing of the compact and non-compact
coordinates makes the separation of the radial and angular variables in the
 equation of motion of the minimally coupled scalar field very difficult. 
This in turn makes the understanding of the dual CFT from the gravity point 
of view very difficult.

The equation of motion to  minimally coupled massive  scalar field  
\be
\Box\Phi-m^2\Phi=0
\ee

Upon expanding with $h=0$, we can re-write it as
\bea
&&-[g^{+-}\omega^2+g^{--}M^2+2g^{+-}M\omega+{\bf k}^2+m^2]\Phi+\f{1}{\sqrt{-g}}
\p_r[\sqrt{-g}g^{rr}\p_r\Phi]+\nn &&\f{1}{\sqrt{-g}}
\p_{\theta_a}[\sqrt{-g}g^{\theta_a\theta_b}\p_{\theta_b}\Phi]+
i[g^{+\theta_a}\omega \p_{\theta_a}\Phi+g^{-\theta_a}M \p_{\theta_a}\Phi]=0,
\eea

where $\Phi=\Phi(r,\theta_a) e^{-i\omega x^+-iM x^-+ik_ix_i}$ and $i,j$ 
represent the non-compact spatial directions along the brane world volume 
whereas
$\theta_a$ represent the angular directions, perpendicular to the brane
world volume and note the appearance of $i$ in the last term.

Now if either of the metric components $g^{-\theta_a}$ and $g^{+\theta_a}$
are non-zero and non-trivial functions of the radial coordinate then it 
would be difficult to solve the equation using separation
of variables and hence reading out the dimension of the operator dual to 
scalar field becomes very difficult. However, as an example, for 
non-relativistic D3 brane solution as presented in section $2.3$,
 only $g^{-\chi}$ is non zero and is equal to $-4\sigma$ and hence the 
equations can be solved using separation of variables. In this case, by 
writing $\Phi(r,\theta_a)=\phi(r)Y(\theta_a)$ makes the radial and angular 
parts of the field to obey 
\bea
&&-[4M^2+\f{2}{r^2}M\omega+{\bf k}^2+m^2]\phi+\f{1}{\sqrt{-g}}
\p_r\bigg(\sqrt{-g}g^{rr}\p_r\phi \bigg)=\natural\phi,\nn
&& \f{1}{\sqrt{-g}}
\p_{\theta_a}\bigg(\sqrt{-g}g^{\theta_a\theta_b}\p_{\theta_b}Y \bigg)-
i 4M \sigma \p_{\chi}Y=-\natural Y, 
\eea
where $\natural$ is a constant and 
solving the radial part of the field $\phi$, suggests the operator dual to 
massive scalar field with dimension $\Delta$ obeys 
\be
\Delta(\Delta-4)=\natural+m^2,
\ee
with the solutions $\Delta_{\pm}=2\pm\sqrt{4+\natural+m^2}$, which are exactly 
the dimension of the operator
dual to a massive scalar field as studied in \cite{kw} for $AdS_5$ but with the square of the mass is $\natural+m^2$ instead of $m^2$. 
Analogous to \cite{kw}, we 
can simply read out the BF bound \cite{bf}, which is $\natural+m^2 >-4$. 

The solutions for Dp branes are singular at $r=0$ as the solutions are not
geodesically complete \cite{hr} even though the curvature invariants are 
smooth and we expect that the near extremal 
solution will cloak the singularity behind the horizon. 

\section{Conclusion}

In this paper we have presented  non-relativistic but supersymmetric 
solutions to various Dp branes. The solutions generically preserves one 
quarter supersymmetry and the extra conditions on the Killing spinor is
\be
\Gamma^+\epsilon=0.
\ee

The non-relativistic solution is  determined by doing a supersymmetry 
preserving deformation to the relativistic solution. Generically the 
structure that does this is of the mixing of the light cone directions $dx^+$
and that of the compact direction denoted by 1-form  $C$.  The symmetries 
preserved by the non-relativistic Dp branes, $p\neq 3$, are that of space 
and time  translations, boosts and rotations.

It would be very interesting to understand the effect of adding such a 
term $Cdx^+$ to metric from the dual field theory point of view, especially 
on the superpotential. 

The non-relativistic extremal solutions are singular at $r=0$ as in the 
relativistic case and this deformation do not take that away. In the case of 
solutions constructed on the deformed conifold, the solution is not any more 
singular as the warp factor do not depends on the parameter $\sigma$. However,
it would be interesting to understand the nature of tidal forces at the 
origin following \cite{hr}.

\section{Acknowledgment}
I would like  to thank Aristos Donos and Jerome Gauntlett for early 
participation and my special thanks to Jerome Gauntlett for several useful
 suggestions. 
It is a great pleasure to thank the String theory group 
[Rajesh Gopakumar, Dileep Jatkar, Satchidananda Nayak, Sudhakar Panda,  
Ashoke Sen, Students and Postdocs], HRI, Allahabad for 
their warm company, where a part of the work is done, also would like to 
thank the theory division SINP, Kolkata for support. I would also like to 
thank Cobi Sonnenschein for making some critical comments about the manuscript,
 R. Gopakumar, D. Jatkar and Shibaji Roy for some useful discussions.

\section{Appendix: The spin-connections}
In this section, we explicitly write down the expression to spin-connections that is used in the main text to check the supersymmetry preserved by the solution.

\subsection{For D1 branes}

The spin connections that follows from the tangent space 1-forms are
\bea\label{spin_connections_d1brane}
\omega^{+ r}&=&-\f{3f'}{8f^{\f{9}{8}}} e^+,~
\omega^{- r}=-\f{3f'}{8f^{\f{9}{8}}} e^-+\f{y'}{rf^{\f{5}{8}}}e^{\chi},~
\omega^{- \chi}=-\f{y'}{rf^{\f{5}{8}}} e^r,~
\omega^{ r \chi}=-\f{y'}{rf^{\f{5}{8}}} e^+-X e^{\chi},\nn
\omega^{- \mu}&=&\f{2y}{r^2f^{\f{5}{8}}} e^{\lambda},~
\omega^{\mu\lambda}=-\f{2y}{r^2f^{\f{5}{8}}} e^+-\f{e^{\chi}}{rf^{\f{1}{8}}}-
2\f{cot2\mu}{rf^{\f{1}{8}}}e^{\lambda},~
\omega^{- \lambda}=-\f{2y}{r^2f^{\f{5}{8}}} e^{\mu},~
\omega^{- \alpha}=\f{2y}{r^2f^{\f{5}{8}}} e^{\sigma_3},\nn
\omega^{\alpha\sigma_3}&=&-\f{2y}{r^2f^{\f{5}{8}}} e^+-2
\f{cot2\alpha}{s_{\mu}rf^{\f{1}{8}}}e^{\sigma_3}-\f{cot\mu}{rf^{\f{1}{8}}} 
e^{\lambda}-\f{e^{\chi}}{rf^{\f{1}{8}}},~
\omega^{- \sigma_3}=-\f{2y}{r^2f^{\f{5}{8}}} e^{\alpha},~
\omega^{- \sigma_1}=-\f{2y}{r^2f^{\f{5}{8}}} e^{\sigma_2},\nn
\omega^{- \sigma_2}&=&\f{2y}{r^2f^{\f{5}{8}}} e^{\sigma_1},~
\omega^{\sigma_1 \sigma_2}=\f{2y}{r^2f^{\f{5}{8}}} e^++
\bigg(\f{cot\alpha}{s_{\mu}rf^{\f{1}{8}}}-\f{4}{s_{\mu}s_{2\alpha}rf^{\f{1}{8}}} \bigg)e^{\sigma_3}+\f{cot\mu}{rf^{\f{1}{8}}}e^{\lambda}+ 
\f{ e^{\chi}}{rf^{\f{1}{8}}},\nn
\omega^{\alpha r}&=&X  e^{\alpha},~
\omega^{\alpha\mu}=\f{cot\mu}{rf^{\f{1}{8}}} e^{\alpha},~
\omega^{\sigma_1 r}=X  e^{\sigma_1},~
\omega^{ \sigma_1 \mu}=\f{cot\mu}{rf^{\f{1}{8}}} e^{\sigma_1},~
\omega^{\sigma_1\alpha}=\f{cot\alpha}{rf^{\f{1}{8}}s_{\mu}} e^{\sigma_1},\nn
\omega^{\sigma_1\sigma_3}&=&\f{cot\alpha}{rf^{\f{1}{8}}s_{\mu}} e^{\sigma_2},~
\omega^{\sigma_2\sigma_3}=-\f{cot\alpha}{rf^{\f{1}{8}}s_{\mu}} e^{\sigma_1},~
\omega^{\sigma_2 r}=X  e^{\sigma_2},~ 
\omega^{\sigma_2 \mu}=\f{cot\mu}{rf^{\f{1}{8}}} e^{\sigma_2},~
\omega^{\sigma_2 \alpha}=\f{cot\alpha}{rf^{\f{1}{8}}s_{\mu}} e^{\sigma_2},\nn
\omega^{\sigma_3 r}&=&X  e^{\sigma_3},~
\omega^{\sigma_3\mu}=\f{cot\mu}{rf^{\f{1}{8}}} e^{\sigma_3},~
\omega^{\lambda r}=X  e^{\lambda},~
\omega^{\alpha\lambda}=-\f{cot\mu}{rf^{\f{1}{8}}} e^{\sigma_3},~
\omega^{\sigma_3\lambda}=\f{cot\mu}{rf^{\f{1}{8}}} e^{\alpha},\nn
\omega^{\sigma_1\lambda}&=&\f{cot\mu}{rf^{\f{1}{8}}} e^{\sigma_2},~
\omega^{\sigma_2\lambda}=-\f{cot\mu}{rf^{\f{1}{8}}} e^{\sigma_1},~
\omega^{\mu\chi}=-\f{1}{rf^{\f{1}{8}}} e^{\lambda},~
\omega^{\lambda\chi}=\f{1}{rf^{\f{1}{8}}} e^{\mu},~
\omega^{\alpha\chi}=-\f{1}{rf^{\f{1}{8}}} e^{\sigma_3},\nn
\omega^{\sigma_3\chi}&=&\f{1}{rf^{\f{1}{8}}} e^{\alpha},~
\omega^{\sigma_1\chi}=\f{1}{rf^{\f{1}{8}}} e^{\sigma_2},~
\omega^{\sigma_2\chi}=-\f{1}{rf^{\f{1}{8}}} e^{\sigma_1},~
\omega^{\mu r}=X  e^{\mu}
\eea

where $X= \f{8f+rf'}{8rf^{\f{9}{8}}}$ and the tangent space 1-forms are 
\bea
e^+&=&f^{-\f{3}{8}}dx^+,~e^-=f^{-\f{3}{8}}(dx^-+C),~
e^{r}=f^{\f{1}{8}}dr,~e^{\mu}=rf^{\f{1}{8}}d\mu,~
e^{\alpha}=r f^{\f{1}{8}}s_{\mu}d\alpha,\nn
e^{\sigma_1}&=&r f^{\f{1}{8}}\f{s_{\mu}s_{\alpha}}{2}\sigma_1,~
e^{\sigma_2}=r f^{\f{1}{8}}\f{s_{\mu}s_{\alpha}}{2}\sigma_2,~
e^{\sigma_3}=r f^{\f{1}{8}}\f{s_{\mu}s_{2\alpha}}{4}\sigma_3,~
e^{\lambda}=r f^{\f{1}{8}}\f{s_{2\mu}}{4}[d\lambda+s^2_{\alpha}\sigma_3],\nn
e^{\chi}&=& f^{\f{1}{8}}\f{r}{2}
[d\chi+s^2_{\mu}(d\lambda+s^2_{\alpha}\sigma_3],~
\eea

\subsection{For D2 branes}

The spin connections are 
\bea\label{spin_connections_d2brane}
\omega^{+ r}&=&-\f{5f'}{16f^{\f{19}{16}}}e^+,~
\omega^{- r}=-\f{5f'}{16f^{\f{19}{16}}}e^{-}+\f{y'}{rf^{\f{11}{16}}} e^{\chi},~
\omega^{-\chi}=-\f{y'}{rf^{\f{11}{16}}}e^r,~ \omega^{ r \psi}=-X  e^{\psi},\nn
\omega^{\mu\psi}&=&-\f{2y}{r^2f^{\f{11}{16}}}e^+-
\f{2cot~2\mu}{rf^{\f{3}{16}}}e^{\psi}-\f{1}{rf^{\f{3}{16}}} e^{\chi},
\omega^{-\mu}=\f{2y}{r^2f^{\f{11}{16}}} e^{\psi},~
\omega^{- \psi}=-\f{2y}{r^2f^{\f{11}{16}}}e^{\mu},\nn
\omega^{\theta\phi}&=&\f{2y}{r^2f^{\f{11}{16}}}e^++\f{cot\mu}{rf^{\f{3}{16}}}
e^{\psi}+\f{1}{rf^{\f{3}{16}}} e^{\chi}-
\f{2cot\theta}{s_{\mu}rf^{\f{3}{16}}}e^{\phi},~
\omega^{- \phi}=\f{2y}{r^2f^{\f{11}{16}}}e^{\theta},
\omega^{- \theta}=-\f{2y}{r^2f^{\f{11}{16}}}e^{\phi},\nn
\omega^{\theta\psi}&=&\f{cot\mu}{rf^{\f{3}{16}}} e^{\phi},
\omega^{\phi\psi}=-\f{cot\mu}{rf^{\f{3}{16}}} e^{\theta},~
\omega^{\mu\chi}=-\f{1}{rf^{\f{3}{16}}} e^{\psi},
\omega^{\psi\chi}=\f{1}{rf^{\f{3}{16}}} e^{\mu},~
\omega^{\theta\chi}=\f{1}{rf^{\f{3}{16}}} e^{\phi},\nn
\omega^{\phi\chi}&=&-\f{1}{rf^{\f{3}{16}}} e^{\theta},~
\omega^{r w}=-\f{3f'}{16f^{\f{19}{16}}} e^w,~
\omega^{x_2 r}=-\f{5f'}{16f^{\f{19}{16}}} e^{x_2},~
\omega^{r\mu}=-X e^{\mu},~\omega^{r \theta}=-X e^{\theta},\nn
\omega^{\mu\theta}&=&-\f{cot\mu}{rf^{\f{3}{16}}} e^{\theta},~
\omega^{ r \phi}=-X  e^{\phi},~
\omega^{\mu\phi}=-\f{cot\mu}{rf^{\f{3}{16}}} e^{\phi},~
\omega^{r \chi}=-X e^{\chi}-\f{y'}{rf^{\f{11}{16}}} e^+
\eea
 where the tangent space  1-forms  are defined as 
\bea
e^+&=&f^{-\f{5}{16}}dx^+,~e^-=f^{-\f{5}{16}}(dx^-+C),~
e^{x_2}=f^{-\f{5}{16}}dx_2,~e^r=f^{\f{3}{16}}dr,~
e^{\mu}=f^{\f{3}{16}}rd\mu,\nn 
e^{\theta}&=&f^{\f{3}{16}}\f{rs_{\mu}}{2}d\theta,
~,e^{\phi}=f^{\f{3}{16}}\f{rs_{\mu}s_{\theta}}{2}d\phi,~
e^{\psi}=f^{\f{3}{16}}\f{rs_{2\mu}}{4}\sigma_3,
~e^{\chi}=f^{\f{3}{16}}\f{r}{2}[d\chi+s^2_{\mu}\sigma_3],\nn
e^w&=&f^{\f{3}{16}}dw.
\eea

\subsection{For D3 branes}

In order to proceed, the spin connections are 
\bea\label{spin_connections_d3brane}
\omega^{r\chi}&=&-\f{y'}{rf^{\f{3}{4}}}e^+-X e^{\chi},~
\omega^{- r}=\f{y'}{rf^{\f{3}{4}}}e^{\chi}-\f{f'}{rf^{\f{5}{4}}} e^{-},~
\omega^{-\chi}=-\f{y'}{rf^{\f{3}{4}}}e^r,\nn
\omega^{\mu\psi}&=&-\f{2y}{r^2f^{\f{3}{4}}}e^+-
\f{2cot~2\mu}{rf^{\f{1}{4}}}e^{\psi}-\f{1}{rf^{\f{1}{4}}} e^{\chi},
\omega^{-\mu}=\f{2y}{r^2f^{\f{3}{4}}} e^{\psi},~
\omega^{- \psi}=-\f{2y}{r^2f^{\f{3}{4}}}e^{\mu},\nn
\omega^{\theta\phi}&=&\f{2y}{r^2f^{\f{3}{4}}}e^++\f{cot\mu}{rf^{\f{1}{4}}}
e^{\psi}+\f{1}{rf^{\f{1}{4}}} e^{\chi}-
\f{2cot\theta}{s_{\mu}rf^{\f{1}{4}}}e^{\phi},~
\omega^{- \phi}=\f{2y}{r^2f^{\f{3}{4}}}e^{\theta},
\omega^{- \theta}=-\f{2y}{r^2f^{\f{3}{4}}}e^{\phi},\nn
\omega^{\theta\psi}&=&\f{cot\mu}{rf^{\f{1}{4}}} e^{\phi},
\omega^{\phi\psi}=-\f{cot\mu}{rf^{\f{1}{4}}} e^{\theta},~
\omega^{\mu\chi}=-\f{1}{rf^{\f{1}{4}}} e^{\psi},
\omega^{\psi\chi}=\f{1}{rf^{\f{1}{4}}} e^{\mu},~
\omega^{\theta\chi}=\f{1}{rf^{\f{1}{4}}} e^{\phi},\nn
\omega^{\phi\chi}&=&-\f{1}{rf^{\f{1}{4}}} e^{\theta},~
\omega^{+r}=-\f{f'}{4f^{\f{5}{4}}}e^+,~
\omega^{x_i r}=-\f{f'}{4f^{\f{5}{4}}} e^{x_i},~
\omega^{r\mu}=-X e^{\mu},~\omega^{r \theta}=-X e^{\theta},\nn
\omega^{\mu\theta}&=&-\f{cot\mu}{rf^{\f{1}{4}}} e^{\theta},~
\omega^{ r \phi}=-X  e^{\phi},~
\omega^{\mu\phi}=-\f{cot\mu}{rf^{\f{1}{4}}} e^{\phi},~
\omega^{ r \psi}=-X  e^{\psi},
\eea 
where $X=\f{4f+rf'}{4rf^{\f{5}{4}}}$, $i=2,3$ and 
$e^{\psi}\equiv e^{\sigma_3}$. The 1-forms in tangent space are defined
\bea
e^+&=&f^{-\f{1}{4}}dx^+,~e^-=f^{-\f{1}{4}}(dx^-+C)~, 
e^{x_i}=f^{-\f{1}{4}}dx_{i},~e^r=f^{\f{1}{4}}dr,
~e^{\mu}=f^{\f{1}{4}}rd\mu,\nn
e^{\theta}&=&f^{\f{1}{4}}\f{r s_{\mu}}{2}d\theta,~
e^{\phi}=f^{\f{1}{4}}\f{r s_{\mu}s_{\theta}}{2}d\phi,~
e^{\sigma_3}=f^{\f{1}{4}}\f{r s_{2\mu}}{4}d\sigma_3,~
e^{\chi}=f^{\f{1}{4}}\f{r}{2}[d\chi+s^2_{\mu}\sigma_3]
\eea

\subsection{For D4 branes}

For this case the spin connections are
\bea\label{spin_connections_d4brane}
\omega^{+r}&=&-\f{3f'}{16f^{\f{21}{16}}}e^+,~
\omega^{-r}=\f{ y'}{rf^{\f{13}{16}}}e^{\psi}-\f{3f'}{16f^{\f{21}{16}}}e^-,~
\omega^{-\theta}=-\f{2  y}{r^2f^{\f{13}{16}}}e^{\phi},
\omega^{-\phi}=\f{2 y}{r^2f^{\f{13}{16}}}e^{\theta},\nn
\omega^{-\psi}&=&-\f{ y'}{rf^{\f{13}{16}}}e^{r},
\omega^{x_ir}=-\f{3f'}{16f^{\f{21}{16}}}e^{x_i},
\omega^{\theta\phi}=-\f{2}{r}\f{cot\theta}{f^{\f{5}{16}}}e^{\phi}+
\f{1}{rf^{\f{5}{16}}}e^{\psi}+\f{2 y}{r^2f^{\f{13}{16}}}e^{+},\nn
\omega^{\theta\psi}&=&\f{1}{rf^{\f{5}{16}}}e^{\phi},~
\omega^{\theta r}=Xe^{\theta},~
\omega^{\phi\psi}=-\f{1}{rf^{\f{5}{16}}}e^{\theta},~
\omega^{\phi r}=Xe^{\phi},~
\omega^{\psi r}=Xe^{\psi}+\f{ y'}{rf^{\f{13}{16}}}e^{+},\nn 
\omega^{w r}&=&\f{5f'}{16f^{\f{21}{16}}}e^w,
\eea
where $X=\f{5rf'+16f}{16rf^{\f{21}{16}}}$. The tangent space 1-forms are 
\bea
e^+&=&f^{-\f{3}{16}}dx^+,~e^+=f^{-\f{3}{16}}(dx^-+C),~
e^{x_i}=f^{-\f{3}{16}}dx_i,~
e^r=f^{\f{5}{16}}dr,\nn e^{\theta}&=&f^{\f{5}{16}}\f{r}{2}d\theta,~
e^{\phi}=f^{\f{5}{16}}\f{r s_{\theta}}{2}d\phi,~
e^{\psi}=f^{\f{5}{16}}\f{r}{2}[d\psi+c_{\theta}d\phi],~e^w=f^{\f{5}{16}}dw.
\eea

\subsection{For NS5 branes}

The spin connections $\omega^{ab}$ are 
\bea\label{spin_connections_ns5brane}
\omega^{+r}&=&-\f{f'}{8f^{\f{11}{8}}}e^+,~
\omega^{-r}=\f{ y'}{rf^{\f{7}{8}}}e^{\psi}-\f{f'}{8f^{\f{11}{8}}}e^-,~
\omega^{-\theta}=-\f{2 y}{r^2f^{\f{7}{8}}}e^{\phi},
\omega^{-\phi}=\f{2 y}{r^2f^{\f{7}{8}}}e^{\theta},\nn
\omega^{-\psi}&=&-\f{ y'}{rf^{\f{7}{8}}}e^{r},
\omega^{x_ir}=-\f{f'}{8f^{\f{11}{8}}}e^{x_i},
\omega^{\theta\phi}=-\f{2}{r}\f{cot\theta}{f^{\f{3}{8}}}e^{\phi}+
\f{1}{rf^{\f{3}{8}}}e^{\psi}+\f{2 y}{r^2f^{\f{7}{8}}}e^{+},\nn
\omega^{\theta\psi}&=&\f{1}{rf^{\f{3}{8}}}e^{\phi},~
\omega^{\theta r}=Xe^{\theta},~
\omega^{\phi\psi}=-\f{1}{rf^{\f{3}{8}}}e^{\theta},~
\omega^{\phi r}=Xe^{\phi},~
\omega^{\psi r}=Xe^{\psi}+\f{ y'}{rf^{\f{7}{8}}}e^{+},
\eea
 
where $X=\f{8f+3rf'}{8rf^{\f{11}{8}}},~i=2,3,4,5$ and the tangent space 
1-forms are
\bea
e^+&=&f^{-\f{1}{8}}dx^+,~e^+=f^{-\f{1}{8}}(dx^-+C),~e^{x_i}=f^{-\f{1}{8}}dx_i,~
e^r=f^{\f{3}{8}}dr,\nn e^{\theta}&=&f^{\f{3}{8}}\f{r}{2}d\theta,~
e^{\phi}=f^{\f{3}{8}}\f{r s_{\theta}}{2}d\phi,~
e^{\psi}=f^{\f{3}{8}}\f{r}{2}[d\psi+c_{\theta}d\phi].
\eea

\section{Sphere metrics}
In this section we shall write down the metric of the spheres--$S^3,~~~S^5$ and $S^7$, in terms of the complex coordinates so as to write down the 1-form $C$ in a very simple form.

\subsection{Three Sphere: $S^3$}
The metric of unit radius $S^3$ is 
\be
d\Omega^2_3=\f{1}{4}[\sigma^2_1+\sigma^2_2+\sigma^2_3],
\ee
where the $\sigma_i$'s are defined in eq(\ref{sigma_1_forms}). Let us write the flat 4-space as $ds^2_4=dr^2+r^2d\Omega^2_3$ and also introduce the following complex coordinates
\be
z_1=s_{\f{\theta}{2}} e^{\f{i}{2}(\psi-\phi)},~~~z_2=c_{\f{\theta}{2}} e^{\f{i}{2}(\psi+\phi)}.
\ee
such that $\sum^2_1z_i{\bar z}_i=1$. The 1-form $\sigma_3$ can be expressed in terms of the complex coordinates  as 
\be
\sigma_3=d\psi+c_{\theta}d\phi=Re\bigg(2i[z_1d{\bar z}_1+z_2d{\bar z}_2]\bigg).
\ee

It means the 1-form $C$ written in eq(\ref{ns5_sol}) or eq(\ref{d5_sol}), can be re-written as 
\be\label{C_complex_coordinate_s3}
C=Re\bigg(2i\sigma r^2[z_1d{\bar z}_1+z_2d{\bar z}_2]\bigg),
\ee
where $Re$ is the real part.

\subsection{Five Sphere: $S^5$}
The metric of unit radius $S^5$ is 
\be
d\Omega^2_5=d\mu^2
+
\f{1}{4}s^2_{\mu}(\sigma^2_1+\sigma^2_2+c^2_{\mu}\sigma^2_3)
+\f{1}{4}(d\chi+s^2_{\mu}\sigma_3)^2.
\ee

Let us introduce the following complex coordinates
\be
z_1=s_{\mu}c_{\f{\theta}{2}} e^{\f{i}{2}(\chi+\psi+\phi)},~~~z_2=s_{\mu}s_{\f{\theta}{2}} e^{\f{i}{2}(\chi+\psi-\phi)},~~~z_3=c_{\mu}e^{\f{i}{2}\chi},
\ee
and then the 1-form, 
$C$, of eq(\ref{d3_sol}) can be re-written as 
\be\label{C_complex_coordinate_s5}
C=Re\bigg(2i\sigma r^2 [z_1d{\bar z}_1+z_2d{\bar z}_2+z_3d{\bar z}_3]\bigg).
\ee

\subsection{Seven Sphere: $S^7$}
The metric of unit radius $S^7$ is 
\be
d\Omega^2_7=d\mu^2+ 
  s^2_{\mu}d\alpha^2+
  \f{1}{4}s^2_{\mu}s^2_{\alpha}(\sigma^2_1+\sigma^2_2+c^2_{\alpha}\sigma^2_3)
+\f{1}{4} s^2_{\mu}c^2_{\mu}(d\lambda+s^2_{\alpha}\sigma_3)^2+\f{1}{4}
 (d\chi+s^2_{\mu}(d\lambda+s^2_{\alpha}\sigma_3))^2.
\ee

Let us introduce the following complex coordinates
\be
z_1=s_{\mu}s_{\alpha}c_{\f{\theta}{2}} e^{\f{i}{2}(\lambda+\chi+\psi+\phi)},~~~z_2=s_{\mu}s_{\alpha}s_{\f{\theta}{2}} e^{\f{i}{2}(\lambda+\chi+\psi-\phi)},~~~z_3=s_{\mu}c_{\alpha}e^{\f{i}{2}(\lambda+\chi)},~~~z_4=c_{\mu}e^{\f{i}{2}\chi},
\ee
and then the 1-form, 
$C$, of eq(\ref{d1_sol}) can be re-written as 
\be\label{C_complex_coordinate_s7}
C=Re\bigg(2i\sigma r^2 [z_1d{\bar z}_1+z_2d{\bar z}_2+z_3d{\bar z}_3+z_4d{\bar z}_4]\bigg).
\ee

\subsection{1-form $C$ in Cartesian coordinates }

Let us introduce the complex Cartesian coordinate $Z_j=r z_j=x_j+i y_j$ for all $j$, as an example $j=1$ and $2$ for flat four space, for which the flat space is $ds^2_n=\sum _i dZ_j d{\bar Z}_j=dr^2+r^2 d\Omega^2_{n-1}=\sum _j[dx^2_j+dy^2_j]$ and $n=4,6,8$.

Then the 1-form, $C$ is 
\be\label{C_complex_coordinate}
C=i \sigma \sum _j[Z_jd{\bar Z}_j-{\bar Z}_j dZ_j]=2\sigma\sum_j(x_jdy_j-y_jdx_j).
\ee 

Note that under the transformation: $Z_j\rightarrow e^{i\beta^{(j)}} Z_j$, i.e. different rotations in different planes, in order to make the 1-form $C$ to remain unchanged gives the condition that  the $\beta^{(j)}$  should better be constants.


\begin{thebibliography}{99}

\bibitem{jm} J. M. Maldacena, {\it The large N limit of superconformal 
field theories and supergravity}, Adv. Theor. Math. Phys. {\bf 2} (1998) 231,
hep-th/9711200.
\bibitem{gkp} S. S. Gubser, I. R. Klebanov and A. M. Polyakov, {\it Gauge 
theory correlators from noncritical string theory,} Phys. Lett. {\bf B428}
(1998) 105, hep-th/9802109.
\bibitem{ew}E. Witten, {\it Anti-de Sitter space and holography}, Adv. Theor. 
Math. Phys. {\bf 2} (1998) 253, hep-th/9802150.
\bibitem{agmoo}O. Aharony, S. S. Gubser, J. Maldacena H. Ooguri and Y. Oz, 
{\it Large N field theories, String theory and gravity}, Phys. Rept, {\bf 323}
(2000) 183-386, hep-th/ 9905111.
\bibitem{oa} O. Aharony, {\it The non-AdS/non-CFT correspondence, or three 
different paths to QCD}, hep-th/0212193.
\bibitem{son}D. T. Son, {\it Toward an AdS/cold atoms correspondence: a
geometric realization of the Schr$\ddot{o}$dinger symmetry},
arXiv:0804.3972[hep-th].
\bibitem{bm}K. Balasubramanian and J. McGreevy, {\it Gravity duals for
non-relativistic CFTs}, arXiv:0804.4053[hep-th].
\bibitem{hrr}C. P. Herzog, M. Rangamani and S. F. Ross, {\it Heating up 
Galilean holography}, arXiv:0807.1099[hep-th].
\bibitem{mmt} J. Maldacena, D. Martelli and Y. Tachikawa, {\it Comments on
string theory backgrounds with non-relativistic conformal symmetry},
arXiv:0807.1100 [hep-th].
\bibitem{abm}A. Adams, K. Balasubramanian and J. McGreevy, {\it Hot spacetimes
for cold atoms}, arXiv:0807.1111[hep-th].
\bibitem{hkss}C. P. Herzog, P. Kovtun, S. Sachdev and D. T. Son, {\it Quantum critical transport, duality and M-theory}, Phys. Rev. {\bf D 75} (2007), 085020,
hep-th/0701036.
\bibitem{hkms} S. A. Hartnoll, P. K. Kovtun, M. Muller and S. Sachdev, 
{\it Theory of the Nernst effect near quantum phase transitions in Condensed
matter, and in dyonic black holes}, arXiv: 0706.3215[hep-th].
\bibitem{hy}S. A. Hartnoll and K. Yoshida, {\it Families of IIB duals for 
relativistic CFTs}, arXiv:0810.0298[hep-th].
\bibitem{dg} A. Donos and J. P. Gauntlett, {\it Supersymmetric solutions for
non-relativistic holography}, arXiv: 0901.0818 [hep-th].
\bibitem{klm}S. Kachru, X. Liu and M. Mulligan, {\it Gravity duals of
Lifshitz-like fixed points}, arXiv:0808.1725[hep-th].
\bibitem{ssp}S. S. Pal, {\it Towards Gravity solutions of AdS/CMT},
arXiv:0808.3232[hep-th].
\bibitem{ssp1}S. S. Pal, {\it More Gravity solutions of AdS/CMT},
arXiv:0809.1756[hep-th].
\bibitem{ssp2}S. S. Pal, {\it Anisotropic  Gravity solutions in AdS/CMT},
arXiv:0901.0599[hep-th].
\bibitem{mt} M. Taylor, {\it Non-relativistic holography}, 
arXiv:0812.0530[hep-th].
\bibitem{vw} A. Volovich and C. Wen, {\it Correlation functions in 
non-relativistic holography}, arXiv: 0903.2455 [hep-th].
\bibitem{amsv}A. Adams, A. Maloney, A. Sinha and S. E. Vazquez, {\it 1/N 
Effects in nonrelativistic gauge-gravity duality}, arXiv: 0812.0166[hep-th].
\bibitem{dt} U. H. Danielsson and L. Thorlacius, arXiv: 0812.5088 [hep-th].
\bibitem{ghhlr} E. G. Gimon, A. Hashimoto, V. E. Hubney, O. Lunin and 
M. Rangamani, {\it Black Strings In Asymptotically Plane Wave Geometries}, 
JHEP 0308:035 (2003), hep-th/0306131.
\bibitem{ag} M. Alishahiha and O. J. Ganor, {\it Twisted Backgrounds, Pp Waves And Nonlocal Field Theories}, JHEP 0303:006 (2003), hep-th/0301080.
\bibitem{ssp3} S. S. Pal, {\it Null Melvin Twist to Sakai-Sugimoto
model}, arXiv:0808.3042[hep-th].
\bibitem{mot} L. Mazzucato, Y. Oz and S. Theisen, {\it Non-relativistic branes}, arXiv:0810.3673[hep-th].
\bibitem{ms} M. Schvellinger, {\it Kerr-Ads black holes and non-relativistic 
conformal QM theories in diverse dimensions}, arXiv: 0810.3011 [hep-th].
\bibitem{bk}N. Bobev and A. Kundu, {\it Deformations of holographic duals
to non-relativistic CFTs}, arXiv: 0904.2873 [hep-th].
\bibitem{gca} 
 A.~Bagchi and R.~Gopakumar,
  arXiv:0902.1385 [hep-th].
  $\bullet$ M.~Alishahiha, R.~Fareghbal, A.~E.~Mosaffa and S.~Rouhani,
  arXiv:0902.3916 [hep-th].
$\bullet$
  K.~M.~Lee, S.~Lee and S.~Lee,
  arXiv:0902.3857 [hep-th].
$\bullet$
  A.~Galajinsky and I.~Masterov,
  arXiv:0902.2910 [hep-th].
$\bullet$
  Y.~Nakayama,
  arXiv:0902.2267 [hep-th].
$\bullet$
  Y.~Nakayama, M.~Sakaguchi and K.~Yoshida,
  arXiv:0902.2204 [hep-th].
$\bullet$
  A.~Akhavan, M.~Alishahiha, A.~Davody and A.~Vahedi,
  arXiv:0902.0276 [hep-th].
$\bullet$
  P.~Horava,
  arXiv:0901.3775 [hep-th].
$\bullet$
  M.~Alishahiha and A.~Ghodsi,
  arXiv:0901.3431 [hep-th].
$\bullet$
  A.~Akhavan, M.~Alishahiha, A.~Davody and A.~Vahedi,
  arXiv:0811.3067 [hep-th].
$\bullet$
  Y.~Nakayama, S.~Ryu, M.~Sakaguchi and K.~Yoshida,
  JHEP {\bf 0901}, 006 (2009)
  [arXiv:0811.2461 [hep-th]].
$\bullet$
  M.~Rangamani, S.~F.~Ross, D.~T.~Son and E.~G.~Thompson,
  JHEP {\bf 0901}, 075 (2009)
  [arXiv:0811.2049 [hep-th]].
$\bullet$
  F.~L.~Lin and S.~Y.~Wu,
  arXiv:0810.0227 [hep-th].
$\bullet$
  C.~Duval, M.~Hassaine and P.~A.~Horvathy,
  arXiv:0809.3128 [hep-th].
$\bullet$
  P.~Kovtun and D.~Nickel,
  Phys.\ Rev.\ Lett.\  {\bf 102}, 011602 (2009)
  [arXiv:0809.2020 [hep-th]].
$\bullet$
  A.~V.~Galajinsky,
  Phys.\ Rev.\  D {\bf 78}, 087701 (2008)
  [arXiv:0808.1553 [hep-th]].
$\bullet$
  D.~Minic and M.~Pleimling,
  arXiv:0807.3665 [cond-mat.stat-mech].
$\bullet$
  J.~W.~Chen and W.~Y.~Wen,
  arXiv:0808.0399 [hep-th].
$\bullet$
  Y.~Nakayama,
  JHEP {\bf 0810}, 083 (2008)
  [arXiv:0807.3344 [hep-th]].
$\bullet$
  W.~Y.~Wen,
  arXiv:0807.0633 [hep-th].
$\bullet$
  M.~Sakaguchi and K.~Yoshida,
  JHEP {\bf 0808}, 049 (2008)
  [arXiv:0806.3612 [hep-th]].
$\bullet$
  J.~L.~B.~Barbon and C.~A.~Fuertes,
  JHEP {\bf 0809}, 030 (2008)
  [arXiv:0806.3244 [hep-th]].
$\bullet$  W.~D.~Goldberger,
  arXiv:0806.2867 [hep-th].
$\bullet$  M.~Sakaguchi and K.~Yoshida,
 ``Super Schrodinger in Super Conformal,''
  arXiv:0805.2661 [hep-th].
$\bullet$ J. P. Gauntlett, S. Kim, O. Varela and D. Waldram, 
arXiv: 0901.0676 [hep-th]. $\bullet$ J. Kluson, arXiv: 0904.1343 [hep-th].
$\bullet$ C. A. Fuertes and S. Moroz,  arXiv: 0903.1844[hep-th].
$\bullet$ I. Gordeli and  P. Koroteev, arXiv: 0904.0509 [hep-th].
$\bullet$ E. O. Colgain and  H. Yavartanoo, arXiv: 0904. 0588 [hep-th].
$\bullet$ F. Correa, V. Jakubsky,  and M. S. Plyushchay, 
Annals Phys.324:1078-1094,2009, arXiv:0809.2854[hep-th].
$\bullet$ P. D. Alvarez, J. L. Cortes, P. A. Horvathy, and M. S. Plyushchay.
JHEP 0903:034,2009, arXiv:0901.1021[hep-th].$\bullet$ A. Bagchi and I. Mandal,
arXiv: 0903.4524 [hep-th]. $\bullet$ D. Martelli and Y. Tachikawa, 
arXiv: 0903.5184 [hep-th].

 

\bibitem{da} D. Anselmi, arXiv:0707.2480, 0801.1216, 0808.3470, 0808.3474.

\bibitem{imsy} N. Itzhaki, J. M. Maldacena, J. Sonnenschein and 
S. Yankielowicz, {\it Supergravity and the large N limit of theories with 
sixteen supercharges}, Phys. Rev. {\bf D 58} (1998) 046004, 
arXiv:hep-th/9802042. 
\bibitem{cd}P. Candelas and X. de la Ossa, {\it Comments on conifold}, 
Nucl. Phys. {\bf B 342} (1990) 246.
\bibitem{ghk} S. S. Gubser, C. P. Herzog and I. R. Klebanov, {\it Symmetry
breaking and axionic strings in the warped deformed conifold}, 
arXiv:hep-th/0405282.
\bibitem{kt} I. R. Klebanov and A. A. Tseytlin, {\it Gravity duals of supersymmetric $SU(N)\times SU(N+M)$ gauge theories}, hep-th/0002159.
\bibitem{ks} I. R. Klebanov and M. Strassler, {\it  Supergravity and a Confining Gauge Theory: Duality Cascades and $\chi$SB-Resolution of Naked 
Singularities}, JHEP {\bf 0008} (2000) 052, hep-th/0007191. 
\bibitem{hr} G. T. Horowitz and S. F. Ross, hep-th/9704058, hep-th/9709050.
\bibitem{kw} I. R. Klebanov and E. Witten, {\it AdS/CFT correspondence and
symmetry breaking}, Nucl. Phys. {\bf B 556}, (1999) 89,
hep-th/9905104.
\bibitem{bf}P. Breitenlohner and D. Z. Freedman, {\it Stability in gauged extended supergravity},
Ann. Phys. {\bf  144} (1982) 249; {\it Positive energy in anti-de Sitter backgrounds and gauged extended supergravity}, Phys. Lett. {\bf B 115} (1982) 197.
\end{thebibliography}
\end{document}